\documentstyle[preprint,aps,epsf]{revtex}

\def \A {{\bf A}}

\def \D {{\bf D}}

\def \e {{\rm e}}
\def\k{{\bf k}}
\def\q{{\bf q}}
\def\na{\bbox{\nabla}}

\begin{document}
\input{psfig.sty}

\title{Nonperturbative Renormalization and the QCD Vacuum}

\author{Adam P. Szczepaniak$^{1}$ and Eric S. Swanson$^{2}$}

\address{$^1$   Physics Department and Nuclear Theory Center \\
   Indiana University,     Bloomington, Indiana   47405-4202}
\address{$^2$ 
Department of Physics and Astronomy, University of Pittsburgh,
Pittsburgh PA 15260 and \\  Jefferson Lab, 12000 Jefferson Ave,
Newport News, VA 23606.}

\maketitle

\begin{abstract}
We present a self consistent approach to Coulomb gauge Hamiltonian QCD 
which allows one  to relate single gluon spectral properties to the long 
range behavior of the confining interaction. 
 Nonperturbative renormalization is discussed. The numerical 
 results are in good agreement with phenomenological and
lattice forms of the static potential.

\end{abstract}
\date{Dec, 1999}
\narrowtext

\section{Introduction} 

There is a long history \cite{ad,kane,amer,fm} of using mean field
 techniques  to  explore dynamical chiral symmetry breaking in Quantum 
 Chromodynamics. Chiral symmetry breaking is of fundamental importance
 because it is directly related to the vacuum
 structure of QCD. It also has important practical consequences, such as making
chiral perturbation theory a viable approximation to low energy QCD. 
Mean field equations are typically derived by analyzing a
 truncation  of the Schwinger-Dyson equations or, when dealing with a 
 Hamiltonian,  by performing a Bogoliubov-Valatin transformation on
 the particle basis.

An added complexity with Hamiltonian approaches has to do with renormalization.
Unlike covariant approaches to field theory which generate a finite number of
counterterms, the Hamiltonian formalism necessarily involves a noncovariant truncation
of the theory and hence may generate noncanonical counterterms. As a result 
it may not be possible to remove the ultraviolet cutoff. In this sense, an
effective field theory approach to Hamiltonian-based renormalization becomes
natural.

 The problem of renormalization in models derived from a QCD
 Hamiltonian  is often ignored. An 
 exception was the work of  Adler and Davis \cite{ad} who have noted
 that, when using the BCS Ansatz, consistency with the Ward identities 
 imposes a definite counterterm structure on the gap equation. They also
noted that 
  calculations should be performed with the full Hamiltonian (as
 opposed to the normal ordered Hamiltonian).
 
Although the quark sector of the vacuum  has been studied relatively well,
the gluonic sector has largely been ignored. However, the gluonic vacuum
is also relevant to studies of 
 chiral symmetry breaking because the quark and gluonic gap equations are
coupled and, more importantly, the  gluon propagator is related to the
quark-antiquark confining interaction.  Furthermore, the
 gluonic vacuum determines 
 the properties of the gluonic 
 quasiparticles which in turn may be used to construct a Fock space expansion 
 of hadrons such as glueballs or hybrids\cite{ssjc}.

In this paper we derive and renormalize the gap equation for the purely
gluonic sector of QCD. This work extends three previous papers. The 
first of these  presented the unrenormalized gluonic gap equation and 
 calculated the gluon condensate and glueball spectrum\cite{ssjc}. 
 The other two\cite{ss3,rssjc} described how to renormalize Hamiltonian QCD
using the similarity or flow equation methods for evolving the
renormalization scale. Both methods are necessarily perturbative, which means
that gap equations may be derived with one (or more) loop corrections. 
However, since the phenomena being studied are nonperturbative, it 
is preferable to construct a nonperturbative renormalization scheme from 
the start.  Unfortunately, this means abandoning the elegant 
flow equation methodology which is so well suited to Hamiltonian-based
perturbation theory.

We further extend the previous work by showing how to obtain the long range 
 interaction between color sources. This is done in the BCS Ansatz by 
  simultaneously considering the single gluon spectral 
 function and the non-Abelian Coulomb potential. This subject was
 previously  considered in 
 Ref.~\cite{Swift} and ~\cite{Zwan} (we discuss the various 
  approaches below). 
 We derive a self consistent gap equation for 
the gluon spectral function.  Self consistency arises because the
 non-Abelian potential 
 appearing in this equation depends on the gluon spectral function.
The solution to the gap equation gives rise to an infrared enhancement in
the effective non-Abelian potential which may be identified with linear
confinement.

In the following section we give
 a brief review of QCD in Coulomb gauge and present our regularization
scheme. We then discuss the renormalization scheme adopted, 
 the gap equation, and the effective interaction emerging from the 
 self consistent solution. 
We conclude and discuss future applications in Section III.

\section{QCD in Coulomb Gauge}
Understanding the properties of soft gluons is one of the major challenges 
of hadronic physics. 
It is natural to study soft gluons using the Hamiltonian formulation
of QCD in a physical gauge such as the Coulomb gauge. 
This has many advantages: it is closest in spirit
to quantum mechanical models of QCD, all degrees of freedom are physical,
no additional constraints are need on the Fock space, the norm is positive
definite, spurious retardation effects are minimized, and confinement
may be rigorously identified with the non-Abelian Coulomb interaction
in the heavy quark limit \cite{ss2}. Finally this is also the natural choice 
for the study of nonrelativistic
 bound states of a few constituent degrees of freedom. It is   
 of relevance in the gluonic sector since one would expect 
 the gluonic quasiparticles to be heavy due to 
   strong confining interactions. 

  The pure gauge QCD Hamiltonian in Coulomb gauge may be written 
as\cite{cl} $H = H_g + H_C$,
where the terms are given by,

\begin{eqnarray}
H_g &=& {\rm Tr}\int d^3x\thinspace \left[{\cal J}^{-1} {\bf \Pi} \cdot
{\cal J} {\bf \Pi} + {\bf B}\cdot{\bf B} \right] \\
H_C &=&  {1 \over 2} g^2 \int d^3x d^3y\thinspace
{\cal J}^{-1}\rho^a({\bf x}) K^{ab}({\bf x},{\bf y}) {\cal J} \rho^b({\bf y})
\end{eqnarray}

\noindent
The second term is the instantaneous non-Abelian Coulomb interaction
for QCD. The kernel $K$ is given by
\begin{equation}
K^{ab}({\bf x},{\bf y}) = \langle{\bf x},a|(\nabla\cdot {\bf D})^{-1}(-\nabla^2)
(\nabla\cdot{\bf D})^{-1}|{\bf y},b\rangle \label{ck}
\end{equation}

\noindent
and the color charge density $\rho^a$ is given by
\begin{equation}
\rho^a({\bf x}) = f^{abc} {\bf A}^b({\bf x}) \cdot {\bf \Pi}^c({\bf x}).
\end{equation}

\noindent
When expanded in powers of the strong coupling, the Coulomb kernel
contains infinitely many terms arising from the inverse of the adjoint 
covariant derivative:

\begin{equation}
{\bf D}^{ab} = \delta^{ab} \nabla - g f^{abc} {\bf A}^c \cdot \nabla.
\end{equation}

\noindent
At lowest order in the coupling the kernel is given by $K^{ab}({\bf
  x}, {\bf y}) = \delta^{ab}/(4 \pi \vert {\bf x} - {\bf y} \vert )$
  as in QED. 
The Faddeev-Popov determinant appears in the kinetic energy and
Coulomb interaction terms due to the curvature of the gauge manifold
and is given by $ {\cal J} = {\rm det}\left[\nabla\cdot{\bf D}\right]$.
Lastly, the components of the non-Abelian magnetic field are given by

\begin{equation}
B_i^a = \epsilon_{ijk}\left( \nabla_j A^a_k + {g\over2}f^{abc} A^b_j A^c_k
\right).
\end{equation}
It should be stressed that this is a bare Hamiltonian. Regularization leads to 
 counterterms which in  practice  will  depend on the approximation
 schemes  used to diagonalize the Hamiltonian. 
As stated above, our goal is to study the QCD vacuum with the aid of
the BCS Ansatz. 
In the past we have chosen to regulate the Hamiltonian by restricting
its  matrix elements in the basis of eigenstates of the bare 
 kinetic energy to a band-diagonal form \cite{ss3,rssjc}. 
 Although an elegant formulation of the renormalization procedure is 
possible with this choice of basis, it gives rise to lengthy
expressions because
all calculations must be performed in the particle basis. Furthermore, 
 such a scheme does not seem to be particularly relevant when dealing with 
nonperturbative renormalization since there is no particular advantage
in using the  bare basis. 
  Hence, it is preferable to employ the field basis and we introduce a 
 field-based regulator  which is analogous to Schwinger's point splitting.
This consists of smearing field operators over a small spatial region:

\begin{equation}
\tilde{\rm A}^b_i({\bf x}) = \int d^3 y \, {\Lambda^3\over (2 \pi)^{3/2}} 
{\rm A}^b_i({\bf y})
{\rm e}^{-({\bf x} - {\bf y})^2 {\Lambda^2\over 2}}
\end{equation}

\noindent
Here $\Lambda$ is the UV cutoff and the fields are effectively smeared
over a distance $O(1/\Lambda)$. 
The mode expansions are 

\begin{eqnarray}
\tilde{\rm A}^b_i({\bf x}) &=& \int {d^3k \over (2 \pi)^3} \,{1 \over \sqrt{2 \omega_0(k)}}
\left( a_i^b({\bf k}) + 
a_i^{b \dagger}(-{\bf k}) \right) {\rm e}^{-i {\bf k}\cdot {\bf x} - {k^2\over 2\Lambda^2}} \\
\tilde\Pi^b_i({\bf x}) &=& \int {d^3k \over (2 \pi)^3} \,i \sqrt{\omega_0(k) \over 2}
\left( a_i^b({\bf k}) - 
a_i^{b \dagger}(-{\bf k}) \right) {\rm e}^{-i {\bf k}\cdot {\bf x} -
{k^2 \over 2 \Lambda^2}}  \label{mode}
\end{eqnarray}

\noindent
where $\omega_0(k) = \vert {\bf k} \vert$ in the perturbative vacuum. We
shall subsequently drop the tildes.

\noindent Contractions of the field operators which are needed below are given as
follows

\begin{eqnarray}
\langle {\rm A}^a_i({\bf x}) {\rm A}^b_j({\bf y}) \rangle &=&  \delta^{ab} \int {d^3 k \over
(2 \pi)^3}\, {D_{ij}(k) \over 2 \omega_0(k)} {\rm e}^{i {\bf k} \cdot ({\bf x}
- {\bf y}) - {k^2 \over \Lambda^2}} \\
\langle \Pi^a_i({\bf x}) \Pi^b_j({\bf y}) \rangle &=&  \delta^{ab} \int {d^3 k \over
(2 \pi)^3}\, D_{ij}(k) {\omega_0(k) \over 2} {\rm e}^{i {\bf k} \cdot ({\bf x}
- {\bf y}) - {k^2 \over \Lambda^2}} \\
\langle {\rm A}^a_i({\bf x}) \Pi^b_j({\bf y}) \rangle &=&  i\delta^{ab} \int {d^3 k \over
(2 \pi)^3}\, {D_{ij}(k) \over 2} {\rm e}^{i {\bf k} \cdot ({\bf x}
- {\bf y}) - {k^2 \over \Lambda^2}}
\end{eqnarray}
where $D_{ij}(k) = \delta_{ij} - \hat k_i \hat k_j$ is the transverse delta
function associated with Coulomb gauge.

\subsection{Nonperturbative Renormalization}

In order to reduce the cutoff dependence induced by regularization 
we allow the couplings to be $\Lambda$-dependent and add an infinite
set of counterterms to the Hamiltonian. The counterterms are 
organized in powers of the cutoff,

\begin{equation}
\delta H \equiv \sum_{n=-2} {c_n(\Lambda)\over \Lambda^n} {\cal O}^{(n)}.
\end{equation}

\noindent
As long at the cutoff only affects the operator products at short
relative distances, 
 the  counterterms ${\cal O}$  must be local. Thus they may be classified 
according to their canonical dimension, $n+1$. 
 Furthermore they have to preserve the unbroken symmetries {\it i.e.} be 
 rotationally invariant, color, spin, and flavor singlets. 
Notice that the series starts at order $\Lambda^2$. This is because
the lowest dimension 
operator satisfying the above restrictions is 
$\int d{\bf x} {\bf A}^2({\bf x})$ 
which is of dimension $-1$, ($n=-2$). The needed marginal 
 and relevant operators contributing to the gluon sector of the 
Hamiltonian are 

\begin{eqnarray}
\delta H &=&  {M^2(\Lambda) \over 2} \int d^3x A^2 
 + { c_0(\Lambda) \over 2}  \int d^3x  A \nabla^2 A
\nonumber \\
&& + {Z^{-1}(\Lambda) - 1\over 2} \int d^3x  \Pi^2 
 +  {Z(\Lambda) - 1\over 2} \int d^3x  {\bf B}^2 
\end{eqnarray}

\noindent
We have established contact with more traditional  notation  by
defining $M^2(\Lambda) = \Lambda^2 c_{-2}$ where $c_{-2}$ is
dimensionless.
 Operators of dimension three and five do not occur. Thus 
the next set of relevant operators are of dimension six:
$f_{abc}f_{cde} A^{ia} \Pi^{ib} A^{jd}\Pi^{je}$,
$A\cdot \Pi A\cdot\Pi$, $A \nabla^4 A$, $\Pi \nabla^2 \Pi$, $f_{abc} \Pi^{ia} 
A^b \cdot \nabla \Pi^{ic}$, combinations of six gluon fields and gradients, 
and so on.

It is important to recall that in an effective field theory the cutoff is not removed. 
 Rather the various coefficients are
 determined by requiring that observables are accurate to a given
 order in  $p/\Lambda$, where $p$ is some 
characteristic, measurable momentum scale. 
In our case the 
   $c_n(\Lambda)$
 should be tuned to reproduce experimental data calculated in some 
nonperturbative 
 scheme. 
   Here we are referring to a number of standard many-body calculational
schemes such as the BCS vacuum Ansatz, the Tamm-Dancoff truncation, or
the random phase approximation.  

Transverse gluons do not contribute to the counterterms because
we employ a BCS Ansatz in this work.
 Thus at this stage of the calculation, the 
 $\Lambda$ dependence of the counterterms will not match that of
 perturbative QCD. 
 Nevertheless couplings to the transverse 
 gluons with momenta below the cutoff are still present in the Hamiltonian and 
lead to important effects which (if the valence sector dominates) could be
taken  into account in perturbation theory around 
 quasiparticle bound states. 
 Removal of the cutoff without taking into account the effects from 
transverse gluons would clearly be incorrect. This will be seen
explicitly below when we compare, in the weak coupling limit, 
  the  counterterms calculated with the BCS
Ansatz with the results coming from 
perturbation theory.


\subsection{The Gap Equation}

There are many equivalent formulations of the BCS approach. The one 
appropriate to many-body physics is the Bogoliubov-Valatin (BV) canonical
transformation on the particle operators. In our case this may be written 
as

\begin{eqnarray}
a^b_i({\bf k}) &=& c(k) \alpha^b_i({\bf k}) + s(k) \alpha^{b\dagger}_i(-{\bf k}) \\
a^{b\dagger}_i({\bf k}) &=& c(k) \alpha^{b\dagger}_i({\bf k}) + s(k) \alpha^{b}_i(-{\bf k}) 
\end{eqnarray}

\noindent
where the rotation is parameterized in terms of an unknown gap function, $\omega$, as

\begin{eqnarray}
c(k) &=& {1\over 2} \left( \sqrt{\omega_0(k) \over \omega(k)} + \sqrt{\omega(k)\over \omega_0(k)} \right) \\
s(k) &=& {1\over 2} \left( \sqrt{\omega_0(k) \over \omega(k)} - \sqrt{\omega(k)\over \omega_0(k)} \right).
\end{eqnarray}

\noindent
Because we have employed a field-based regulator, the
effect of the BV transformation is to simply replace $\omega_0(k)=k$ with
$\omega_0=\omega(k)$ in the mode expansions of Eq.~(\ref{mode}). Thus all matrix elements
evaluated in the BCS vacuum are simply related to those evaluated in the
perturbative vacuum with the same replacement. 

The gap equation may now be obtained from the one-body portion of the
QCD Hamiltonian.
Normal ordering with respect to the BCS vacuum yields

\begin{equation}
H_{1b} = \int {d^3 q \over (2 \pi)^3}
  \e^{-{k^2\over\Lambda^2}}
 \left[ E(q)
  \alpha^{b\dagger}_i({\bf q}) \alpha^{b}_i({\bf q}) + G(q) 
\left( \alpha^b_i({\bf q}) \alpha^b_i(-{\bf q}) + H.c. \right) \right]
\end{equation}

\noindent
where

\begin{eqnarray}
E(q) &=& {1\over 2\omega(q)} \Bigg[  
  {\omega^2(q)\over Z} + Z q^2  + M^2(\Lambda)
  + q^2 c_0(\Lambda) 
  \nonumber  \\
&& + {N_c \over 4} \int {d^3 k \over (2 \pi)^3}\,  {\rm e}^{-{k^2\over \Lambda^2}}\,   \tilde V({\bf k}+{\bf q}) \, (1 +
(\hat k \cdot \hat q)^2) \,  {\omega^2(k) + \omega^2(q) \over \omega(k) }  \nonumber \\
&&  + {\pi \alpha_s(\Lambda) N_c } \int {d^3 k \over (2 \pi)^3} \,   {\rm e}^{-{k^2\over \Lambda^2}} \, {(3 - (\hat k \cdot \hat q)^2) \over \omega(k)} 
\Bigg]
\end{eqnarray}

\noindent
and

\begin{eqnarray}
G(q) &=& {1 \over 4\omega(k)} \Bigg[  
 - { \omega^2(q) \over Z} +   Z q^2    + M^2(\Lambda)
 + c_0(\Lambda)q^2  \nonumber \\
&+& {N_c \over 4} \int {d^3k\over (2 \pi)^3} \, 
\tilde V({\bf k} + {\bf q}) \e^{-{k^2\over \Lambda^2}}\, (1 + 
 (\hat k \cdot \hat q)^2)\, {\omega^2(k) - \omega^2(q) \over
   \omega(k)} 
\nonumber \\
&+& {\pi \alpha_s(\Lambda) N_c } \int {d^3k \over (2 \pi)^3} \,
\e^{-{k^2\over \Lambda^2}} 
 { {(3 - (\hat k \cdot \hat q)^2) }\over \omega(k)}
 \Bigg]
\end{eqnarray}
The spectral function $\omega(k)$ is obtained from solving the gap equation
\begin{equation}
G(q) = 0
\end{equation}
which demands that the quasiparticle, BCS vacuum 
decouples from states with pairs of gluons --a feature reminiscent of the 
constituent quark model. 

The static potential in these expressions  is the expectation value 
 of the non-Abelian Coulomb kernel in the BCS vacuum state,

\begin{equation}
\tilde V({\bf q}) = \int d^3 r \,{\rm e}^{-i{\bf q}\cdot{\bf r} } 
 \langle BCS \vert K^{ab}({\bf r},{\bf 0}) \vert BCS \rangle.
\end{equation}

\noindent
At lowest order in the coupling and in the perturbative vacuum this potential
is simply $4 \pi \alpha_s(\Lambda) / q^2$. It is possible to show that the 
non-Abelian Coulomb term describes the complete quark-antiquark interaction in 
the heavy quark limit\cite{ss2}. Thus one may use lattice data to
determine that 
the potential takes on the familiar Coulomb+linear form. 
 It is natural to assume that, with the appropriate color
structure,  this form holds for gluonic 
color sources as well (in fact this is confirmed by lattice calculations
\cite{latt}). Thus, in previous calculations\cite{ss2,rssjc}  we have
simply replaced the expectation of the non-Abelian kernel with a linear
potential. However, one of our current goals is to demonstrate that 
the BCS Ansatz and linear confinement can be described consistently
by the same formalism.
The central idea is that the gap equation contains a kernel which is itself
the expectation value of the non-Abelian Coulomb interaction in the BCS 
vacuum. Thus the kernel is functionally dependent on the gap function and it
is possible to obtain both the gap function and the effective potential.
 Since
it is known that the non-Abelian Coulomb interaction gives rise to 
confinement in the heavy quark limit, one can hope that this
procedure will yield a linear potential at large
distances~\cite{Swift},\cite{Zwan}. 

In an attempt to gain some insight into this hypothesis, 
we consider perturbative corrections to the effective potential.
As stated above, the 
non-Abelian  Coulomb kernel is an operator which 
 depends on the transverse degrees of freedom of the gauge field 
(cf. Eq.~(\ref{ck})). It can be shown that the 
 $\beta$ function can be obtained from loop corrections to this
 kernel \cite{F}.  One finds that at 
  $O(\alpha_s)$, there are three contributions to $\beta = - \beta_0/(2 \pi) \cdot  \alpha_s^2$:
 $\beta_0 = \beta_c + \beta_g + \beta_q = + 4N_c +
 (- 1) + ( - 2n_f/3) $. The first term, $\beta_c$ comes directly from the 
expansion
of the non-Abelian Coulomb potential.
 The second and third terms are due to mixing with
 two-gluon and two-quark 
intermediate states  respectively.  In the following we consider the
perturbative corrections to the Coulomb kernel since it gives rise to 
the majority of the pure gauge $\beta$ function.

The contribution from the transverse gluons 
in the expansion of the Coulomb kernel may be computed with the expansion

\begin{equation}
[{1 \over {\na\cdot\D}}(-\na^2) {1 \over {\na\cdot\D}}]_{ab} = 
-{1\over \na^2} \left[ \delta_{ab} + 2 g_1 {1\over \na^2} f_{abc}\A^c\cdot \na + 
3 g_2 {1\over \na^2} f_{acd} \A^d \cdot \na {1\over \na^2} f_{cbe} \A^e \cdot \na + \ldots \right]
\end{equation}
with  the cutoff dependence of the couplings, $g_i \equiv g (\Lambda) Z_i(\Lambda)$
to be determined. 
 To order $g_i^2$, the  Coulomb 
interaction is given by 

\begin{equation}
V_c(q) = {4 \pi\alpha_s\over q^2 } \left[ 1 + {3 
    Z^2_2(\Lambda) \alpha_s(\Lambda) 
\beta_c \over 16 \pi^2 } 
I(q,\Lambda;\omega_0) \right]
\end{equation}

\noindent
where $\alpha_s(\Lambda)=g^2(\Lambda)/4\pi$ and 

\begin{equation}
I(q,\Lambda;\omega_0) = \int {d^3 {\bf k}} {(1-(\hat \k \cdot \hat{\bf q})^2)\,
{\rm e}^{-k^2/\Lambda^2} \over  \omega_0(k) ({\bf q}-\k)^2}.
\end{equation}

In the BCS vacuum the same expression arises except
that $\omega_0$ is replaced with the unknown function, $\omega$. Thus,
as stated above, 
the solution to the gap equation depends on a potential which itself is a
function of the gluon dispersion relation. It therefore becomes possible
to obtain the gluon gap function and the nonperturbative potential in 
a self consistent fashion.

We pursue this scenario by defining a nonperturbative coupling via the
relation

\begin{equation}
V_c(q) \equiv {4 \pi \tilde{\alpha}_{eff}(q^2) \over q^2}.
\end{equation}

\noindent
Furthermore, we take part of the rainbow-ladder approximation to $V_c$
by summing  all  diagrams proportional to $I^n$. 
This leads to an expression familiar from the
 perturbative leading log approximation,

\begin{equation}
{\tilde\alpha}_{eff}(q^2) = {\alpha_s(\Lambda)  \over  
 1 - {3Z^2_2(\Lambda)\alpha_s(\Lambda)\beta_c \over {16\pi^2 }} I(q,\Lambda;\omega)}.
\end{equation}

\noindent
Although this potential may be inserted directly into the gap equation we have
found it useful to approximate it with the aid of the following substitution
 
\begin{equation}
I[\omega] \to {4 \pi \over 3}  \log\left[ {{\Lambda^2 }\over {\omega(q)^2}}\right]
\label{approx}.
\end{equation}
As shown in Fig.~1, this form is accurate to roughly
10\% for $q/\Lambda < 0.1$.

\begin{figure}[hbp]
\hbox to \hsize{\hss\psfig{figure=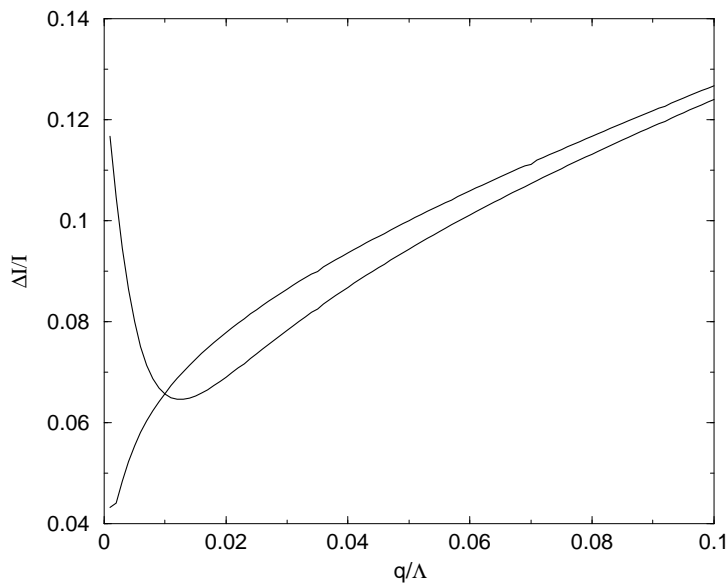,width=3.8in,angle=0}\hss}
\vspace{0.5cm}
{\caption{\label{fig:amb}
 Relative error due to approximation defined in Eq.~(\ref{approx}). The 
ratio $(I[\omega]-\log\left[ {{\Lambda^2 }\over {\omega(q)^2}}\right])/
I[\omega]$ is plotted as a function of $q/\Lambda$ for $\omega(q)^2 \equiv M^2 +
q^2$. The two curves correspond to $M/\Lambda = 0.01$ and
$M/\Lambda=0.1$ (larger difference for smaller $q$). }}
\end{figure}

With this expression for $V_c$, the gap equation is given by

\begin{eqnarray}
{ \omega(q)^2 \over Z} &=& Z q^2  + M^2(\Lambda) + c_0(\Lambda)q^2 
 +
 {\pi \alpha_s(\Lambda) N_c } \int {d^3k \over (2 \pi)^3} \,
\e^{-{k^2\over \Lambda^2}} 
 { {(3 - (\hat k \cdot \hat q)^2) }\over \omega(k)} \nonumber \\
&+& {N_c \over 4} \int {d^3 k \over (2 \pi)^3} \, \e^{-{k^2\over
 \Lambda^2}}\,\tilde V_c({\bf k} +{\bf q}) \, \left( 1 + (\hat q \cdot
 \hat k)^2 \right) \,
 {\omega^2(k)-\omega^2(q) \over \omega(k)}  \label{baregap}
\end{eqnarray}

\noindent

Many of the terms of higher order in $1/\Lambda$ also contribute to the
gap equation. Additional terms include $c_2(\Lambda) q^2  \omega(q)^2/\Lambda^2$,
$d_2(\Lambda) q^4/\Lambda^2$, and $e_2(\Lambda)/\Lambda^2  \cdot N_c/3 
\cdot\int  {\rm exp}(-k^2/\Lambda^2) (\omega(k)^2 - \omega(q)^2)/\omega(k)$. 
As expected, these only affect the solution at
short range, thus the long range behavior of the effective potential is
completely specified by the nonperturbative model (the BCS Ansatz in this
case). Nevertheless, one may get an indication about the viability of 
the low energy model by studying the coefficients, $c_2$,
$d_2$, etc -- a strong cutoff dependence at low momenta would indicate that 
the nonperturbative model is inadequate and that additional 
counterterms are therefore  required. 

At this point the standard procedure would be to choose values for the
cutoff and coefficients, 
 solve the gap equation to obtain $\omega$,
 use the solution to obtain, say, glueball masses\footnote{This procedure is
followed, without renormalization, in Ref. \cite{ssjc}.}, and vary the
coefficients so that the predictions agree with experimental (or lattice) data.
However, in this study we choose to perform a simpler analysis and
fit the derived form of the effective potential, $V_{eff} = 
\tilde V(\omega)$ to the Wilson loop lattice potential. This is similar in spirit 
to a lattice renormalization procedure advocated by Lepage and 
Mackenzie\cite{LM}. Dimension six operators are neglected 
so that the couplings to be varied are the five coefficients 
$Z$, $c_{-2}$, $c_0$, $Z_2$, and $\alpha_s$.
First we note that not all of the coefficients are independent since 
one requires 
that the effective potential, and therefore $\omega$,
 is $\Lambda$ independent (to leading order in $p/\Lambda$). 
  Multiplying  the gap equation by $Z$ and introducing 
   $\tilde{\alpha_s} = Z \alpha_s$,
$\tilde c_{-2}  = Z (c_{-2} + L)$, $\tilde c_0 = Z^2 + Z c_0 $ leads to 

\begin{equation}
\omega^2(q) = {\tilde c}_0(\Lambda) q^2 + \Lambda^2 {\tilde c}_{-2}(\Lambda) 
 +  {N_c \over {(4\pi)^2}} \int_0^\infty d k k^2 d(\hat q \cdot \hat k)
 \tilde V_c({\bf k} +{\bf q}) \, \left( 1 + (\hat q \cdot \hat k)^2 \right) \, {\omega^2(k) - 
\omega^2(q) \over \omega(k)}  \label{baregap2}
\end{equation} 
where 
 
\begin{equation}
\tilde V_c(q) = { {4\pi{\tilde \alpha_s}(\Lambda)} \over   {q^2\left(1 -
    {{{\tilde \alpha_s}(\Lambda)\beta_c}\over {4\pi} } \log\left[
    {{\Lambda^2}\over {\omega^2(q)}}\right]\right) } }
\end{equation}
and $L$ is defined to be the constant,
$\Lambda$-dependent contribution from the third term on the right hand side of
Eq.~(\ref{baregap}) (this term comes from the $A^4$ operator in the 
Hamiltonian). In obtaining Eq.~(\ref{baregap2}) we have further chosen $Z = Z^2_2$ 
which is required to make the effective interaction cutoff
independent. 

We now define a QCD scale, $\mu$, as 

\begin{equation}
\mu^2 \equiv \Lambda^2 {\rm e}^{-{4\pi \over {\tilde \alpha}_s(\Lambda) \beta_c}}
\end{equation}

\noindent
and set $m^2(q) \equiv \omega^2(q) -q^2$. The  expression
for the running coupling becomes cutoff independent (as long as $m(q)$
is $\Lambda$-independent) and is given by

\begin{equation}
\alpha_{eff}(q^2) = { 4 \pi \over \beta_c {\rm log}({q^2+m(q)^2 \over
    \mu^2})}
,\;\; {\tilde V}_c(q) =  {{4\pi\alpha_{eff}(q^2)}\over {q^2}} \label{veff}.
\end{equation}

\noindent
We note that if $m(0) = \mu$  the long range behavior of the effective 
 potential corresponds to a linear potential in position space. 
 Equating the coefficient of the leading $1/q^{4}$ behavior   
 with the standard form $6\pi b/q^4$ ($b$ is the string tension),
  yields $\mu^2 = 3 (1+m')\beta_c b/(8 \pi)$ where 
  $m' = dm^2(0)/dq^2$.  The two remaining coefficients in the gap
 equations can thus be fixed by demanding that the effective potential 
  represents linear confinement with the right slope. 
  In numerical computations, instead of fixing $m'$ 
  we have chosen to fix $m^2(q)$ at a finite value of $q^2$, typically 
  $q^2 = 1\mbox{ GeV}^2$, which is numerically easier.  
 The two renormalization conditions, one for $m(0)$ and one for $m'$ (or
 in general for $m^2(q_0)$ and $m^2(q_1)$), 
 can be implemented directly into the gap equation by two subtractions,

\begin{eqnarray}
\omega^2(q) &= & q^2 {{ \omega^2_1-\omega^2_0 }\over {q^2_1 - q^2_0}}
 + {{ q^2_1 \omega^2_0 - q^2_0\omega^2_1 } \over {q^2_1 - q^2_0 }}
 + \int_0^\infty d k {\cal V}(k,q) {\omega^2(k) - \omega^2(q) \over \omega(k)}  \nonumber \\
&  + & {{q^2 - q^2_1}\over {q^2_1-q^2_0}}
 \int_0^\infty d k {\cal V}(k,q_0) {\omega^2(k) - \omega^2_0 \over \omega(k)} 
 + {{q_0^2 - q^2}\over {q^2_1-q^2_0}}
 \int_0^\infty d k {\cal V}(k,q_1) {\omega^2(k) - \omega^2_1 \over \omega(k)} 
\end{eqnarray}
where $\omega_0 = \omega(q_0)$, $\omega_1 = \omega(q_1)$ and 
\begin{equation}
{\cal V}(k,q) = {{N_c}\over {(4\pi)^2}} \int_{-1}^{1} d ({\hat q}\cdot {\hat k})
 k^2 {\rm e}^{-{k^2\over \Lambda^2}} \, {\tilde V_c}(\k+\q)
(1+({\hat q}\cdot {\hat k})^2) 
\end{equation}

\noindent
The original couplings may be reconstructed as follows:

\begin{equation}
\Lambda^2 {\tilde c}_{-2} = {{ q^2_1 \omega^2_0 - q^2_0\omega^2_1 } 
\over {q^2_1 - q^2_0 }} 
-  {{q^2_1}\over {q^2_1-q^2_0}}
\int_0^\infty d k {\cal V}(k,q_0) {\omega^2(k) - \omega^2_0 \over \omega(k)} 
 + {{q^2_0}\over {q^2_1-q^2_0}}
 \int_0^\infty d k {\cal V}(k,q_1) {\omega^2(k) - \omega^2_1 \over \omega(k)} 
\end{equation}

\begin{equation}
{\tilde c}_0  =   {{ \omega_1^2 - \omega_0^2} \over {q^2_1 - q^2_0}}
 - {1\over {q_1^2-q_0^2}}
 \int_0^\infty d k {\cal V}(k,q_1) {\omega^2(k) - \omega^2_1 \over \omega(k)} 
 + 
 {1\over {q_1^2-q_0^2}}
 \int_0^\infty d k {\cal V}(k,q_0) {\omega^2(k) - \omega^2_0 \over \omega(k)}
\end{equation}

\noindent
The single gluon energy is given by 
\begin{equation}
E(q) = \omega(q)\left[ Z^{-1} + \int_0^\infty dk {{\cal V}(k,q)\over
    \omega(k)}\right] \label{ener}
\end{equation}

If a linear potential is to be recovered (as it must because we are fitting
to lattice data), one must have $\omega_0 =  \mu$ at $q_0=0$, 
 (see Eq.~(\ref{veff})). Thus the QCD scale $\mu$ must be related 
 to the string tension. 
  As discussed above, instead of taking the limit $q_1 \to q_0$ and 
fixing $d\omega^2(0)/dq^2$ to reproduce the magnitude of 
string tension  we keep 
$q_1$ finite, choose 
 $q_1 = 1\mbox{ GeV}$ and fit $\omega_1$ to reproduce the strength of the  
confining potential at low momentum. 
In our computations 
the
one dimensional nonlinear integral (after numerical evaluation of the angular integrals) 
equation was solved numerically using a modified Levenberg--Marquardt 
algorithm. We note that the equations are extremely sensitive
to the proper treatment of the IR singularity (which, however, is integrable) 
 and that some solutions which have appeared in the literature
are misleading. The results for $\alpha_{eff}(q) = q^2V_{eff}(q)/4\pi$ are shown in Fig.~2.
We find that $m' \sim 0$ (typically $m'=0.01$) thus the QCD scale $\mu$
can be directly calculated from the physical string tension. Taking
$b=0.18\mbox{ GeV}^2$ yields $\mu^2=0.2579\mbox{ GeV}^2$. Finally, for large momenta 
 the running of the effective coupling is given by

\begin{equation}
\alpha^{UV}_{eff}(q) = {{4\pi}\over {{\beta_c}\log\left( {{q^2}\over
        {0.2579 GeV^2}} \right)}}.
\end{equation}

Over the a range of momenta the full solution to the 
 effective coupling (or potential) 
 can be well approximated by a running coupling version of the 
 Cornell potential
\begin{equation}
\alpha_{eff}(q) =  {{4\pi}\over {\beta_c\log\left[ {{q^2}\over {\mu^2}} +
      \left({{\mu_0}\over {\mu}}\right)^2\right]}}
 + {3\over 2} {b\over {q^2}} = {{\alpha^{UV}_{eff}(\Lambda)}\over {\left(1  
 + {{\alpha^{UV}_{eff}(\Lambda)\beta_c}\over {4\pi}}
\log{ {{p^2 + \mu_0^2}\over {\Lambda^2}} }\right)}}  + {3\over 2} {b\over {q^2}}.
\end{equation}

\noindent
Here $\mu_0$ is a parameter which allows interpolation between the infrared and
ultraviolet limits of the potential. We find $(\mu_0/\mu)^2 \sim 10$. 
Essentially
identical results are obtained upon comparison with the standard Coulomb+linear
form of the Cornell potential.

\begin{figure}[htb]
\hbox to \hsize{\hss\psfig{figure=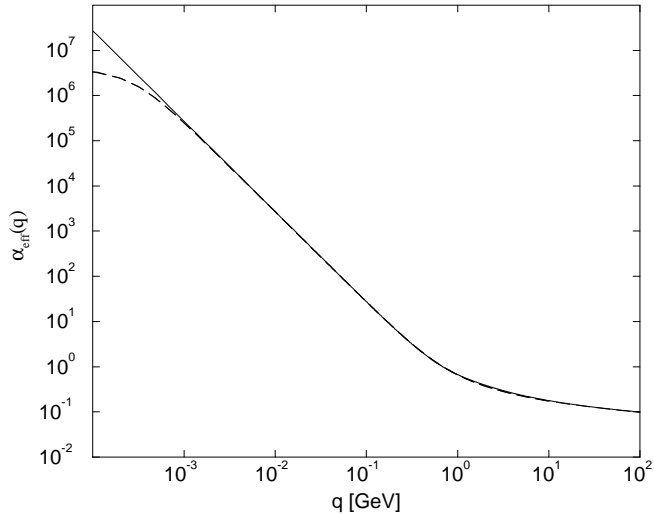,width=3.8in}\hss}
\vspace{0.5cm}
{\caption{
The effective potential derived from the self-consistent gap
equation. The dashed line is the numerical solution
to the gap equation with $\Lambda=10 \mbox{ GeV}$  ($\Lambda$ dependence is negligible and would 
not be visible on this scale).  The solid line is the fit to the running Cornell
potential as discussed in the text. The discrepancy at low momenta is due to the 
finite mesh size in the numerical calculations.  }}
\end{figure}
 That the form of the derived effective potential so nearly reproduces
 linear confinement at long range is a tantalizing indication that our goal
 of deriving the confinement potential in a self-consistent manner from the
 BCS vacuum Ansatz may be achieved. 

The derived gluon dispersion relation $\omega(q)$ is shown in Fig.~3.
The doubly subtracted form of the gap equation  removes $\Lambda^2$ and
log$(\Lambda)$ dependence from the gap equation and leaves subleading
terms of order $q^2/\Lambda^2$. This is reflected in the cutoff independence
evident in Fig.~3 at low momentum.
In Fig.~3 we also compare the full solution of the gap equation
for $\Lambda=10$ with the approximate one obtained using the
``Coulomb+linear'' potential with the parameters fitted to the
self consistent solution as discussed above.  Both of them lead to $m' \sim 0$ 
and both are renormalized at the same point, $q_1=1$ GeV, however the 
 intermediate $q$ dependence is somewhat different. As seen from Fig.~2, 
this small difference does not alter the behavior of the
  effective interaction.

\begin{figure}[htb]
\hbox to \hsize{\hss\psfig{figure=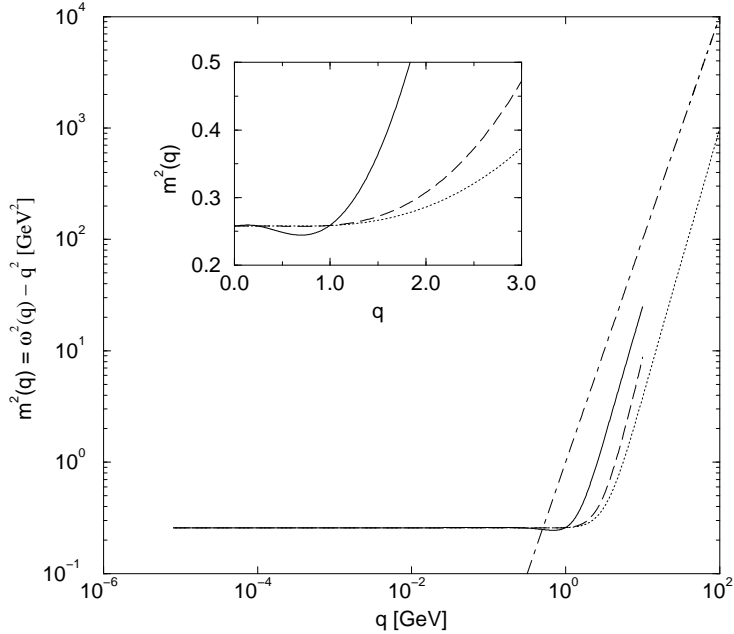,width=3.8in}\hss}
\vspace{0.5cm}
{\caption{ $m^2(q) \equiv \omega^2(q) - q^2$ is a function of momentum. The dashed and dotted lines 
 correspond to $\Lambda=10\mbox{ GeV}$ and $\Lambda=100\mbox{ GeV}$ 
 respectively. The solid line is the 
  Cornell potential fit  to the full solution with 
  $\Lambda=10\mbox{ GeV}$. The dash-dotted line is $m^2(q) = q^2$. }}
\end{figure}

One may expect that $\omega(q)$ should approach the 
 free solution, $\omega_0(q) = q$ at large momenta. However, this need
not be true, both in perturbation theory, where logarithms can ruin the 
asymptotic behavior, and nonperturbatively.  This is because the 
counterterms $Z\, B^2$ and 
$c_0\,A\nabla A$ give rise to the term  ${\tilde c}_0 q^2 $
in the gap equation. 
 Renormalizing $\omega(q)$ at a low momentum scales
 $q_0,q_1 << \Lambda$ will result in a logarithmic dependence on $\Lambda$ in
 ${\tilde c}_0$. Thus the  
 effective gluon mass $m^2(q) = \omega^2(q) - q^2$ will remain
proportional to $q^2$ at large momenta.

As mentioned above, this behavior is also true in perturbation theory. In the
following we examine the small coupling limit of the counterterm coefficients to
establish contact with perturbative QCD.
Evaluating Eqns 37 and 38  to $O\left(\alpha_s={\tilde \alpha}_s(\Lambda)\right)$ yields

\begin{equation}
{\tilde c}_{-2} = {{\alpha_s N_C}\over {4\pi}}\left[ -{4\over 3} -
  {8\over 3} \right] = -{{\alpha_s \beta_0}\over  N_C} \label{tc2}
\end{equation}
 where the first contribution comes from the vacuum expectation value 
 of the Coulomb potential and the second from the $A^4$ term in the 
Hamiltonian, 
 and 
\begin{equation}
{\tilde c}_0 = 1 + {{\alpha_s N_c}\over {4\pi}}{8\over 15}{{q^2_1 + q^2_0}\over {q^2_1-q^2_0}}
\ln\left({q_0\over q_1}\right) +  {{\alpha_s N_c}\over {4\pi}}{8\over
  15} \ln \left({\Lambda^2\over {q_0 q_1}}\right)
 =  {{\alpha_s N_c}\over {4\pi}}{8\over
  15} \ln \Lambda^2 + \mbox{finite}. \label{c0}
\end{equation}
In perturbation theory, to $O(\alpha_s)$, $\Lambda$-dependence of $Z$
 and $c_0$ follows from Eq.~(\ref{c0}) and the requirement that
the gluon energy (an observable in perturbation theory) is independent of
 $\Lambda$. Thus from Eq.~(\ref{ener}) it follows that 
\begin{equation}
E(q) = q\left[ Z^{-1} + {{\alpha_s N_c}\over {4\pi}}{4\over
    3}\ln{{\Lambda^2}\over {q^2}} 
\right]
\end{equation}
which leads to 
\begin{equation}
Z = {{\alpha_s N_c}\over {4\pi}}{4\over
    3}\ln\Lambda^2 + \mbox{finite}
\end{equation}
 and finally
\begin{equation}
c_0 =  -{{\alpha_s N_c}\over {4\pi}}{32\over
    15}\ln\Lambda^2 
\end{equation}

These expressions agree with the corresponding expressions in a full perturbative
Hamiltonian QCD calculation\cite{rssjc}. Note, however, that in  
Ref.~\cite{rssjc} it has been shown that $c_0 = 0$ to 
$O(\alpha_s)$. This arises because the 
 self energy contribution from a gluon loop with transverse gluon exchange 
cancels 
 the contribution from the Coulomb 
 and the $A^4$ vacuum expectation values\footnote{  
 Note that the UV cutoff $\Lambda$ used in Ref.~\cite{rssjc} is by a 
 factor of $\sqrt{2}$ larger then the one used here.}. 
The BCS calculation only contains the latter two terms. 
 In a nonperturbative
 calculation  based on the BCS Ansatz
 transverse gluons are confined and the single gluon energy of
 Eq.~(\ref{ener}) becomes IR singular. As a result, transverse gluons 
only affect observables in bound state calculations of color singlet 
states.

In Fig.4 we show the behavior of the counterterms, ${\tilde c}_{-2}$ and
${\tilde c}_{0}$ for small couplings. The points represent the full expressions 
obtained from Eqns 37 and 38. The lines are the perturbative results given above
with the finite portions fit to the full results. As expected the agreement is very
good for small coupling. Fig. ~4a shows a dramatic dependence of $-\tilde{c_{-2}}$ 
on the value of the
UV cutoff. This dependence originates with the subleading $1/q^4$ behavior of 
the effective potential which makes a log$(\Lambda)$ contribution to
$\tilde c_{-2}$. The solid lines in Fig.~4a incorporate this behavior. The logarithmic
term dominates in the small coupling limit  and leads to the large relative splitting
seen in the figure.

\begin{figure}[htb]
\hbox to \hsize{\hss\psfig{figure=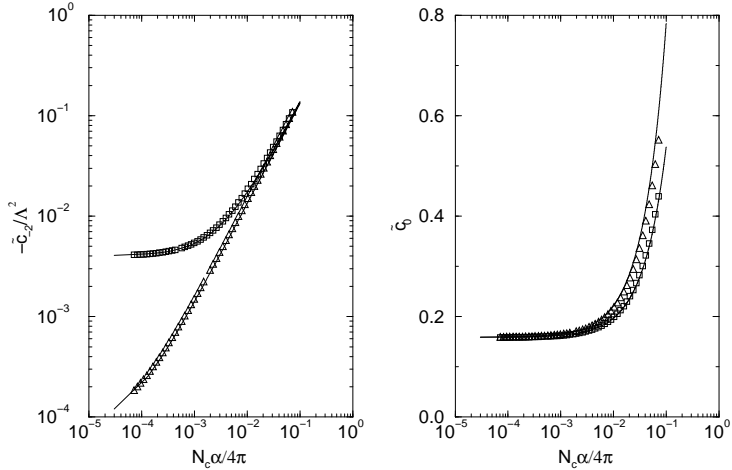,width=3.8in}\hss}
\vspace{0.5cm}
{\caption{ Counterterms as a function $\alpha_s$ in the weak coupling limit. 
  The two set of points correspond to $\Lambda=10\mbox{ GeV}$
   (squares) and $\Lambda=100\mbox{ GeV}$ 
 (triangles). Solid lines are the results of the leading order 
  $O(\alpha_s)$ perturbative approximation given by Eqs.~(\ref{tc2}) and 
   ~(\ref{c0}).  } }
\end{figure}

\subsection{Relationship to Other Approaches}

The Schwinger-Dyson formalism if often used to discuss the issues 
presented here\cite{CP,RW}. Advantages are that it is covariant
and that calculations beyond the rainbow approximation are possible.
However, the resulting integral equations are four dimensional and are
therefore difficult to solve. They are also normally evaluated in Euclidean
space which introduces interpretational difficulties, especially
when dealing with confinement.
Furthermore, 
the connection to the parton model is unclear
since it is difficult to continue wavefunctions to Minkowski space.
 In the Hamiltonian
  framework color nonsinglets decouple from the physical spectrum as in the
Schwinger-Dyson approach, but 
hadronic wavefunctions are well-defined, and thus a parton picture of
high energy QCD is possible.

 We also note that mean field formalisms of 
Coulomb gauge QCD have been extensively studied in Refs.~\cite{Swift}
and \cite{Zwan}. The numerical solution presented in
Ref.~\cite{Swift} are however doubtful, as they do not have the
proper large momentum behavior. In Ref.~\cite{Zwan} the Coulomb
potential was studied with the aid of a model of the gluon dispersion function 
(the gap equation was not solved).

A calculation with similar goals to those expressed here has recently
appeared\cite{gjc}. This paper employs the flow equation renormalization
methodology to derive a gluonic gap equation which is similar in form
to Eq.~(\ref{baregap}).
Differences are that convergence factors have been inserted by hand into 
two of their loop integrals, a phenomenological potential has been 
used for the interaction kernel, and the perturbative expression for the
gluon mass counterterm has been used. Their calculation also includes transverse
gluon contributions at one loop, which is incorrect since the low
momentum transverse gluon propagator is strongly modified by the
confining interaction. As stated above, the implication of this is that any
propagator of color nonsinglet states is IR divergent and that transverse
gluons should only be incorporates into color singlet bound state calculations.
 We stress our belief that the 
interaction can, and should, be derived in a self-consistent fashion and
that the use of perturbative renormalization is not a viable approach to
nonperturbative problems. A related problem is seen in their Figure 2 which
shows $\omega(q) - q$ approaching zero for $q > 4$ GeV. This is not correct,
$\omega$ will in fact {\it not} approach $q$ at large momentum, but some 
multiple of $q$.

Curtis and Pennington \cite{CP} have noted that the 
gap equation derived in the Schwinger-Dyson formalism is not 
multiplicatively renormalizable. This means that it is not possible
to exchange the $\Lambda$-dependence of the gap solution with a 
renormalization point dependence.  As stated above, multiplicative 
renormalizability is not a concern in the effective field theory.
 An obvious manifestation of the difference is the appearance of
 noncanonical counterterms like the gluon mass term. 
 In our approach the cutoff has to be kept finite, otherwise
neglected contributions, {\it i.e.} from transverse gluons,
would cause new divergences to appear. 
Instead we keep 
$\Lambda$ finite and use higher dimensional local operators
to account for short distance
contributions from transverse gluons. In this way, the neglected 
contributions are
kept finite (below the cutoff) and can be evaluated by expanding the
Fock space used to diagonalize the Hamiltonian.

\section{Conclusions}

We have presented a computational scheme for evaluating vacuum and 
quasiparticle properties of QCD. The scheme is based on the QCD Hamiltonian
in Coulomb gauge, a field-based regularization, and the effective field
theory approach to renormalization. The use of the Hamiltonian is 
advantageous because all degrees of freedom are physical and a 
transparent picture of confinement emerges -- confinement is generated
by the instantaneous non-Abelian Coulomb interaction.   We found it 
useful to regulate the theory by field-smearing because this greatly
reduced the complexity of the computations. Field-based
calculations have forced us to adopt a field-based renormalization 
procedure and we have found that the effective field theory approach is
ideal for our work.

A gap equation was derived from the QCD Hamiltonian using the Bogoliubov-Valatin
transformation. The gap equation necessarily involves the Coulomb interaction
kernel which should also be evaluated in the BCS vacuum. Thus self-consistent
gap and effective instantaneous color source interaction equations  may
be derived. With the aid of an approximation to the matrix element of the
full kernel in the BCS vacuum, we have found that it is possible to
obtain an effective potential which agrees well with the (Abelian) Coulomb
and linear regimes of the Wilson loop  measurement of the heavy source 
nonperturbative potential. Furthermore, the derived effective potential
is nearly cutoff independent, indicating that a good low energy description
has been found.

Encouraged by this success, we look forward to repeating the calculation with
the exact BCS interaction kernel.  We fully expect this calculation to be
equally successful. It will also be of interest to examine the consistency
of our results with those derived in the quark sector. Since the quark and gluon vacuum sectors couple at one loop, an examination of the coupled gap equations
should prove illuminating. Finally the efficacy of the methodology presented 
here can be tested by calculating glueball and meson spectra.

\acknowledgements 
ES acknowledges support from the DOE under
grant DE-FG02-96ER40944 and DOE contract DE-AC05-84ER40150 under
which the Southeastern Universities Research Association operates
the Thomas Jefferson National Accelerator Facility. AS acknowledges
support from the DOE under grant DE-FG02-87ER40365.

\end{document}